\documentclass[review,floats,floatfix]{elsarticle}
\usepackage{lineno,latexsym,graphicx}
\usepackage[usenames,dvipsnames]{color}
\usepackage{hyperref}
\usepackage{amsmath}
\usepackage{amssymb}
\usepackage{multirow}
\usepackage[utf8]{inputenc}
\usepackage{array}
\usepackage{booktabs}
\usepackage{tabularx}
\UseRawInputEncoding
\modulolinenumbers[5]
\RequirePackage{ifpdf}
\journal{Computational Materials Science}

\biboptions{sort&compress}

\begin{document}

\begin{frontmatter}

\title{Simulation of the Einstein--de Haas effect combining molecular and spin dynamics}



\author[UNISA,DFA]{W. Dednam\corref{mycorrespondingauthor}}
\cortext[mycorrespondingauthor]{Corresponding author}
\ead{dednaw@unisa.ac.za}

\author[DFA]{C. Sabater}
\author[UNISA]{A. E. Botha}
\author[UNISA]{E. B. Lombardi}
\author[INL]{J. Fern\'andez-Rossier}
\author[DFA]{M. J. Caturla}

\address[UNISA]{Department of Physics, Unisa Science Campus, University of South Africa, Johannesburg 1710, South Africa}
\address[DFA]{Departamento de F\'\i sica Aplicada and Unidad asociada CSIC, Universidad de Alicante, Campus de San Vicente del Raspeig, E-03690 Alicante, Spain.}
\address[INL]{ International Iberian Nanotechnology Laboratory, 4715-310 Braga, Portugal}

\date{\today}

\begin{abstract}
The spin and lattice dynamics of a ferromagnetic nanoparticle are studied via molecular dynamics and with semi-classical spin dynamics simulations where spin and lattice degrees of freedom are coupled via  a dynamic uniaxial anisotropy term. We show that this model conserves total angular momentum, whereas spin and lattice angular momentum are not conserved. We carry out simulatios of the  the Einstein--de Haas effect for a Fe nanocluster with more than 500 atoms that is free to rotate, using a modified version of the open-source spin-lattice dynamics code (\texttt{SPILADY}). We show that the rate of angular momentum transfer between spin and lattice is proportional to the strength of the magnetic anisotropy interaction. The addition of the anisotropy allows full spin-lattice relaxation to be achieved on previously reported timescales of $\sim100$ ps and for  tight-binding magnetic anisotropy energies comparable to those of small Fe nanoclusters.
\end{abstract}


\end{frontmatter}


\section{Introduction}
First observed by Einstein and De Haas~\cite{EdH1915} in 1915, the so-called Einstein--de Haas (EdH) effect involves the rotation of a magnetic object through change in its magnetization. The converse effect, i.e. magnetization through a rotational change, was discovered contemporaneously by Barnett~\cite{Barnett1915}.
Both effects rely on the conservation of the total angular momentum and are governed by the same gyromagnetic tensor components, satisfying the Onsager reciprocal relations~\cite{lan95,bau10,bau12}. Thus, for any particular physical system, experimental confirmation of either effect is thought to be sufficient to establish conservation of total angular momentum~\cite{fuk19}. Arguably, the EdH effect is more widely known because of its routine use in measuring the $g$ factors of various ferromagnetic materials: Fe, Co, Ni, etc.~\cite{Scott1962}. Recently, the EdH effect has also been observed in a ferromagnetic thin film, deposited on a microcantilever~\cite{wal06}. 

Despite several theoretical studies~\cite{kov07,jaa09,chu10,Mentink2019}, a microscopic description of the two effects, either in terms of computer simulations or theoretically, remains a challenging and open problem.
On the theoretical front the most recent work by Mentink~\textit{et al}.~\cite{Mentink2019} shows that, to accurately determine the timescale of angular momentum exchange between electronic and lattice degrees of freedom, and to establish whether the exchange is local or non-local, it is necessary to cast both quantum subsystems in the angular momentum representation. In practice, this involves ``dressing'' every atom with phonons that exchange angular momentum locally with the multi-electron atom's spins and electronic orbital angular momenta. The quantum many-body Hamiltonian obtained in this way is fully rotationally invariant and a static non-perturbative variational calculation shows that phonons quench the electronic orbital angular momentum of the atoms as the coupling strength between the lattice and electronic degrees of freedom increases.  A dynamic non-perturbative variational calculation by the same authors \cite{Mentink2019}, on the other hand, reveals the emergence of high frequency peaks in the magnetic susceptibility as the electron-phonon coupling strength increases. These peaks are consistent with the timescale of the ultra-fast EdH effect that was observed recently in experiments of laser-induced demagnetization in an isolated ferromagnetic sample~\cite{Dornes2019}.

Transfer of angular momentum between spin and motional degrees of freedom ultimately occurs due to spin-orbit interactions that couple these degrees of freedom. In the spin-lattice dynamics (SLD) model, spin-orbit interactions lead to magnetic anisotropy terms that depend explicitly on the atomic positions, thereby coupling spin and lattice degrees of freedom
\cite{Beaujouan2012,PereraThesis2015,Perera2016,AssmannThesis2018,Assmann2018,Hellsvik2018,Tranchida2018,Nieves2020,Strungaru2021}. Importantly, total angular momentum must be conserved. 

In this work we implement the SLD model of Ref. \cite{Perera2016}. The model is based on a truncated Taylor expansion of the magnetic anisotropy energy \cite{Eisenbach2015}. We demonstrate that it is able to reproduce spin-lattice relaxation times $\sim 100$ ps, consistent with previously reported values \cite{Beaujouan2012,Perera2016,Tranchida2018}. Our implementation of the SLD model is applicable to combined molecular and spin dynamics of systems of arbitrary symmetry (including bulk, clusters, defect systems, etc.) It allows the modelling of a wide range of potential new systems at finite temperatures, for example: non-collinear magnetic domain walls in ferromagnetic atomic-sized contacts or wires \cite{Dednam2020Ni,Caridad2014,Fini2014}, chiral spin textures in magnetic nanoparticles \cite{Reeves_2014,Chauleau2020}, skyrmions \cite{Fert2017,Zazvorka2019}, and an atomistic description of the Barnett \cite{Barnett1915} and Barkhausen \cite{Cullity2009} effects, amongst others.

The remaining material is organized as follows. In Section 2, we give a brief outline of the SLD model of ferromagnetic exchange previously implemented in \texttt{SPILADY} \cite{spilady2016} and demonstrate analytically that it conserves angular momentum. However, we show that angular momentum exchange is not possible when only the isotropic ferromagnetic exchange term is considered in the model.  We therefore show analytically in Section 2 that, by adding the second-order anisotropy correction of Ref. \cite{Perera2016} to the SLD model, not only is total angular momentum conserved, but also exchange of angular momenta between spin and lattice is facilitated. In Section 3, we discuss our choice of semi-classical potentials and parameters to perform SLD simulations of ferromagnetic iron nanoclusters and then demonstrate numerically that our implementation replicates the Einstein--de Haas effect in a small iron nanocluster. Finally, we conclude the paper by summarizing our findings in Section 4. 

\section{Models and analytical considerations}

\subsection{General description of the SLD model}

Our implementation builds on the well-established technique known as spin-lattice dynamics (SLD) \cite{Ma2008,Beaujouan2012,spilady2016,Tranchida2018,Assmann2018}. This is a semi-classical molecular dynamics approach which allows the simultaneous evolving of atomic spin and position degrees of freedom. The atom-atom interactions in this model are described by the widely-used embedded atom method (EAM) potential \cite{daw1983semiempirical}. 
The EAM potential has been implemented in \texttt{SPILADY} \cite{Ma2008,spilady2016} along with a generalized Heisenberg model of ferromagnetic exchange:
\begin{linenomath*}
\begin{equation}
H =\sum_{i}^{N}\frac{\overrightarrow{p_{i}}^{2}}{2m}+V+H_{\text spin}\, 
\label{eq1}
\end{equation}
\end{linenomath*}
In Eq.~(\ref{eq1}), $\overrightarrow{{p}_{i}}$ is the momentum operator for the $i$th atom and $V$ is the interatomic EAM potential energy function that models the basic interactions between the atoms, i.e. excluding the spin interactions. Expressed in terms of the spin angular momentum vectors of particles $i$ and $j$, the spin Hamiltonian $H_{\rm spin}$ is a sum of the Zeeman term ($H_{\rm Zeeman}$), the generalized Heisenberg exchange term ($H_{\rm exch}$) and the anisotropy term ($H_{\rm anis}$):

\begin{linenomath*}
\begin{equation}
H_{\rm spin} = H_{\rm Zeeman} + H_{\rm exch} + H_{\rm anis} 
\label{eq1a}
\end{equation}
\end{linenomath*}
with
\begin{linenomath*}
\begin{equation}
H_{\rm Zeeman} =
-\sum_i \vec{M}_i\cdot\vec{B}_{\text ext}=\sum_i g\mu_B \vec{S}_i\cdot\vec{B}_{\text ext}, 
\label{eq2}
\end{equation}
\end{linenomath*} and
\begin{linenomath*}
\begin{equation}
H_{\rm exch} =  -\frac{1}{2}\sum_{i} J_{ij}\left(r_{ij}\right)\left[{\vec{ S}}_i\cdot{\vec{S}}_j - \left|{\vec{ S}}_i\right|\left|{\vec{S}}_j\right| \right].
\label{eq3}    
\end{equation}
\end{linenomath*}
The anisotropy term, $H_{\rm anis}$, is discussed in the next section. The Zeeman term in Eq.~(\ref{eq2}) is due to an external magnetic field $\vec{B}_{\rm ext}$,  where $\vec{M}_i = -g\mu_B \vec{S}_i$ is the magnetic dipole moment, $g=2.002319$ is the electron's gyromagnetic factor and $\mu_{\rm B}$ is the Bohr magneton. The exchange term, Eq. (\ref{eq3}), is isotropic in form, with $r_{ij} = |\vec{r}_i - \vec{r}_j|$ being the magnitude of the displacement $\vec{r}_{ij} = \vec{r}_i - \vec{r}_j$ between the $i$th and $j$th atoms.  ${\vec{S}}_i$ is the (dimensionless) atomic spin or spin angular momentum vector at site $i$ and the quantities $J_{ij}\left(r_{ij}\right)$ are obtained from a shifted fitting function of the form \cite{Ma2008,Perera2016}
\begin{linenomath*}
\begin{equation}
 J_{ij}\left(r_{ij}\right) = J_{0}\left(1-r_{ij}/r_c\right)^3\Theta\left(r_c-r_{ij}\right),
\label{eq4a}   
\end{equation}
\end{linenomath*}
in which $J_0$ is the only fitting parameter, $\Theta\left(r_c-r_{ij}\right)$ is the Heavyside step function, and $r_c$ is the cut-off radius. 

Next, we discuss the conservation of total angular momentum in the Heisenberg model of ferromagnetic exchange implemented in \texttt{SPILADY}.

\subsection{Conservation of total angular momentum in SLD simulations}

We briefly revisit the textbook derivation of angular momentum conservation in a system of classical particles. 
We define the total  angular momentum of a system of classical particles:
\begin{linenomath*}
\begin{equation}
\vec{L}=\sum_i \vec{L}_i
\label{eq5a}
\end{equation}
\end{linenomath*}
where $\vec{L}_i = m_i \vec{r}_i\times\vec{v}_i$
is the angular momentum of an individual particle. 
The rate of change of the total angular momentum is governed by the following equation:
\begin{linenomath*}
\begin{equation}
\frac{d\vec{L}}{dt}=\sum_i \frac{d\vec{L}_i}{dt}=
\sum_i \vec{r}_i\times \vec{F}_i
\label{eq6a}
\end{equation}
\end{linenomath*}

We can break down the force acting on particle $i$ into two parts, the forces coming from external sources outside of the system and the internal forces, 
$\vec{F}_i=\left(\vec{F}_{i,\rm{ext}} + \vec{F}_{i,\rm{int}}\right)$, where the internal forces are given by 
\begin{linenomath*}
\begin{equation}
\vec{F}_{i,\rm{int}}=\sum_{j\neq i} \vec{f}_{ji}
\label{eq7a}
\end{equation}
\end{linenomath*}
and where $\vec{f}_{ji}$ is the force that particle $j$ exerts on $i$. By Newton's third law, $\vec{f}_{ji}=-\vec{f}_{ij}$. Therefore, for any pair of particles $i$ and $j$, the contributions to the rate of change of the angular momentum deriving from their mutual interaction only, are:
\begin{linenomath*}
\begin{equation}
\left. \frac{d\vec{L}_i}{dt} \right|_{{\rm int},j}+ \left.\frac{d\vec{L}_j}{dt}\right|_{{\rm int},i}= 
\vec{r}_i\times  \vec{f}_{ji} + \vec{r}_j \times\vec{f}_{ij}= \left(\vec{r}_i-\vec{r}_j\right)\times\vec{f}_{ji}
\label{eq4}
\end{equation}
\end{linenomath*}

Now, if the internal forces $\vec{f}_{ij}$ are  central, then they do not contribute to the rate of change of angular momentum of the system, and we are left with only the external forces. 
Moreover, a force $\vec{f}_{ij}=-\vec{\nabla}U(|\vec{r}_i-\vec{r}_j|)$ that can be obtained as a gradient of a potential $U$, such as the EAM potential \cite{daw1983semiempirical} used in this work, is automatically central and, by virtue of being expressed as a gradient of a scalar function,  conservative. 

This shows that, for an isolated system of classical particles, in the absence of external forces, and with internal forces that are central, the rotational angular momentum is conserved. 

Equation (\ref{eq2}) shows that the atomic spins in our implementation are dimensionless. Consequently, in order to calculate the dimensionless classical angular momentum, we define $\vec{\lambda} = \frac{\vec L}{\hbar}$ with ${\hbar}=6.58211899\times10^{-16}$ eV.s in the following expressions involving the classical angular momentum. So, substituting the Heisenberg exchange term, Eq. (\ref{eq3}), into Eq. (\ref{eq4}) yields for the rate of change of the lattice angular momentum:

\begin{linenomath*}
\begin{equation}
\begin{gathered}
\left.  \frac{d \vec{\lambda}_i}{dt} \right|_{{\rm exch},j}+ \left. \frac{d \vec{\lambda}_j}{dt}\right|_{{\rm exch},i}=
 \frac 1{\hbar}\left[\vec{r}_i\times  \vec{f}_{ji} + \vec{r}_j \times\vec{f}_{ij}\right]=
 \frac 1{\hbar}\left[\vec{r}_i\times-\vec{\nabla_i}H_{{\rm exch,}j} + \vec{r}_j \times-\vec{\nabla_j}H_{{\rm exch,}i}\right]\\=
-\vec{r}_i\times\frac 1{\hbar r_{\mathit{ij}}}\frac{dJ_{ij} (r_{\mathit{ij}})}{dr_{\mathit{ij}}}\left[{\vec{ S}}_i\cdot{\vec{S}}_j - \left|{\vec{ S}}_i\right|\left|{\vec{S}}_j\right|\right]\vec{r}_{ij} +\vec{r}_j \times\frac 1{\hbar r_{\mathit{ji}}}\frac{dJ_{ji} (r_{\mathit{ji}})}{dr_{\mathit{ji}}}\left[{\vec{ S}}_j\cdot{\vec{S}}_i  - \left|{\vec{ S}}_j\right|\left|{\vec{S}}_i\right| \right]\vec{r}_{ij}\\=
(-\vec{r}_i+\vec{r}_j)\times\frac 1{\hbar r_{\mathit{ij}}}\frac{dJ_{ij} (r_{\mathit{ij}})}{dr_{\mathit{ij}}}\left[{\vec{ S}}_i\cdot{\vec{S}}_j  - \left|{\vec{ S}}_i\right|\left|{\vec{S}}_j\right|\right]\vec{r}_{ij}\\=
\frac 1{\hbar r_{\mathit{ij}}}\frac{dJ_{ij} (r_{\mathit{ij}})}{dr_{\mathit{ij}}}\left[{\vec{ S}}_i\cdot{\vec{S}}_j - \left|{\vec{ S}}_i\right|\left|{\vec{S}}_j\right|\right]\vec{r}_{ij}\times\vec{r}_{ij}=0
\end{gathered}
\end{equation}
\end{linenomath*}

We therefore see that the classical or rotational angular momentum is conserved in an isotropic model of ferromagnetic exchange.

With a view to establishing conservation of spin angular momentum in SLD models, we need to derive an expression analogous to Eq. (\ref{eq4}) involving the rates of change of the angular momentum of spins $i$ and $j$. To that end, we start with the equation of motion for the classical spin momentum of particle $i$:
\begin{linenomath*}
\begin{equation}
\frac{d\vec{S}_i}{dt}=-\vec{S}_i\times\frac{1}{\hbar}\nabla _{\vec{S_i}}H_{\rm spin}
\end{equation}
\end{linenomath*}
where the energy of the spin system, $H_{\rm spin}$, is in general a function of $\vec{S}_i$, $\vec{S}_j$ and of the coordinates of the atoms and their velocities. 

We first compute the effective field associated with the first two terms of Eq. (\ref{eq1a}), classically. For the first term, $H_{\rm Zeeman}$, we obtain:
\begin{linenomath*}
\begin{equation}
\nabla _{\vec{S_i}}H_{\rm Zeeman}=g\mu_B \vec{B}_i
\end{equation} 
\end{linenomath*}
This term produces a rate of change of the spin momentum when $\vec{S}_i$ is not parallel to $\vec{B}_i$, and, consequently, we do not consider total angular momentum conservation in the presence of an external magnetic field any further.

The second term, $H_{\rm exch}$, produces central forces when we compute the associated gradients. Since individual spin magnitudes are conserved in our model, we have for the spin torques:
\begin{linenomath*}
\begin{equation}
\nabla _{\vec{S_i}}H_{\rm exch} = -\frac{1}{2}\sum_j J(r_{ij}) \vec{S}_j 
\end{equation} 
\end{linenomath*}
By analogy with Eq. (\ref{eq4}), we find that for any pair of particles $i$ and $j$ the contribution to the rate of change of the spin angular momentum coming from this term vanishes: 

\begin{linenomath*}
\begin{equation}
\begin{gathered}
\left.\frac{d\vec{S}_i}{dt}\right|_{{\rm exch,}j}+
\left.\frac{d\vec{S}_j}{dt}\right|_{{\rm exch,}i}=\\ 
\frac{J(r_{ij})}{\hbar}\left( \vec{S}_i \times \vec{S}_j + \vec{S}_j\times\vec{S}_i\right)=0
\end{gathered}
\end{equation} 
\end{linenomath*}
In other words, isotropic interactions of the scalar form  $\vec{S}_1\cdot\vec{S}_2$ preserve the spin angular momentum pairwise.   

Therefore, the Heisenberg model of ferromagnetic exchange in Eq. (\ref{eq3}) not only conserves total angular momentum, but does so separately within the spin and lattice sub-reservoirs, and cannot account for angular momentum transfer between lattice and spin degrees of freedom.

In the next section, we consider the correction $H_{\rm anis}$, which couples the spin and lattice degrees of freedom, allowing transfer of angular momentum between spin and lattice, while still conserving total angular momentum.

\subsection{Addition of magnetic anisotropy}

To permit exchange of angular momentum between spins and lattice, we have implemented in the \texttt{SPILADY} code \cite{spilady2016} the model of anisotropy described in Ref. \cite{Perera2016}. We exclude the first-order anisotropy correction of Ref. \cite{Perera2016} since it does not preserve time-reversal symmetry in the absence of an external field. Therefore we only consider the second-order (uniaxial) anisotropy correction of Ref. \cite{Perera2016} in our implementation:

 \begin{linenomath*}
\begin{equation}
H_{\rm anis} = -C_2\sum_{i,\alpha,\beta} S_{\alpha}(i)\Lambda_{\alpha,\beta}(i)S_{\beta}(i),
\label{eq7} 
\end{equation}
\end{linenomath*}
In Eq. (\ref{eq7}), $\Lambda_{\alpha,\beta}(i) = \frac{{ \partial ^{2} \rho_{i}(r_{ij})}} {{\partial x_{i\alpha} \partial x_{ i\beta }}}$, with $\alpha, \beta = 1, 2, 3 $ Cartesian components, is the Hessian matrix element with respect to lattice site $i$, of the semi-classical \textit{many-body}, EAM electron density-like, function:
\begin{linenomath*}
\begin{equation}
\begin{gathered}
\rho_{i}\left(r_{ij}\right) =  \left\{
        \begin{array}{ll}
            \sum_{j \neq i} \left(1-\frac{r_{ij}}{r_0}\right)^{4}e^{\left(1-\frac{r_{ij}}{r_0} \right)} & \quad r_{ij} \leq r_0 \\
            0 & \quad r_{ij} > r_0,
        \end{array}
    \right.
\label{eq8}
\end{gathered}
\end{equation}
\end{linenomath*}
where $r_0$ is the cut-off distance between second and third nearest neighbours and $C_2$ is a proportionality constant which determines the relative strength of the anisotropy term \cite{Perera2016}. \footnote{The interested reader is referred to Ref. \cite{DednamThesis2019} for full details of our implementation in \texttt{SPILADY} of the algorithms described in Refs. \cite{Krech1998,Landau1999,PereraThesis2015,Perera2016}.}

Thus, the contribution of the anisotropy to the effective local magnetic field at atom \textit{i} from neighbouring spins within the cut-off $r_0$ becomes:

\begin{linenomath*}
\begin{equation}
\begin{gathered}\nabla _{\vec{S_i}}H_{\text{anis}}
=-C_2\sum_{j{\neq}i}\left[g\left(r_{\mathit{ij}}\right)\vec{ S_i} +\frac
1{r_{\mathit{ij}}}\frac{\mathit{dg}\left(r_{\mathit{ij}}\right)}{dr_{\mathit{ij}}}\left(\vec{
r}_{\mathit{ij}}\cdot \vec{ S_i}\right)\vec{
r}_{\mathit{ij}}\right]\end{gathered}
\label{eq9}
\end{equation}
\end{linenomath*}

where $g(r_{\mathit{ij}})\text =\frac 1{r_{\mathit{ij}}}\frac{d\phi (r_{\mathit{ij}})}{dr_{\mathit{ij}}}$ \ with 
$r_{\mathit{ij}}\text =|\vec{r_i}-\vec{r_j}|$\ and $\phi (r_{\mathit{ij}})$ is the expression under the summation in Eq. (\ref{eq8}). 

Evaluating $\frac{d\vec{S}_i}{dt}=-\vec{S}_i\times\frac{1}{\hbar}\nabla _{\vec{S_i}}H_{\rm anis}$ for the spin at site $i$ yields
\begin{linenomath*}
\begin{equation}
\begin{gathered}\frac{d\vec{S}_i}{dt}=\frac{C_2}{\hbar}\sum
_{k{\neq}i}\left[\frac
1{r_{\mathit{ik}}}\frac{\mathit{dg}\left(r_{\mathit{ik}}\right)}{dr_{\mathit{ik}}}\left(\vec{
r}_{\mathit{ik}}\cdot \vec{ S_i}\right)\vec{S}_i \times \vec{
r}_{\mathit{ik}}\right]\end{gathered}
\end{equation}
\end{linenomath*}
In the same way, for the spin at site $j$,
\begin{linenomath*}
\begin{equation}
\begin{gathered}\frac{d\vec{S}_j}{dt}=\frac{C_2}{\hbar}\sum
_{k{\neq}j}\left[\frac
1{r_{\mathit{jk}}}\frac{\mathit{dg}\left(r_{\mathit{jk}}\right)}{dr_{\mathit{jk}}}\left(\vec{
r}_{\mathit{jk}}\cdot \vec{ S_j}\right)\vec{S}_j \times\vec{
r}_{\mathit{jk}}\right]\end{gathered}
\end{equation}
\end{linenomath*}

Finally, for the $j$th term from $\frac{d\vec{S}_i}{dt}$ and the $i$th term from $\frac{d\vec{S}_j}{dt}$, summing gives
\begin{linenomath*}
\begin{equation}
\begin{gathered}\left. \frac{d\vec{S}_i}{dt} \right|_{j}+ \left.\frac{d\vec{S}_j}{dt} \right|_{i}
=\frac{C_2}{\hbar r_{\mathit{ij}}} \frac{\mathit{dg}\left(r_{\mathit{ij}}\right)}{dr_{\mathit{ij}}}\left[\left(\vec{
r}_{\mathit{ij}}\cdot \vec{ S_i}\right)\vec{S}_i \times \vec{
r}_{\mathit{ij}}
+ \left(\vec{
r}_{\mathit{ij}}\cdot \vec{ S_j}\right)\vec{S}_j \times\vec{
r}_{\mathit{ij}} \right]
\end{gathered}
\label{eq11}
\end{equation}
\end{linenomath*}
since two instances of $\vec{r}_{\mathit{ji}}$ = -$\vec{r}_{\mathit{ij}}$ occur in the second term so their minus signs cancel when the indices of $i$ and $j$ are swapped.

Next, we consider the lattice torque term, showing that it exactly compensates the spin torque term of Eq. (\ref{eq11}). 

The expression for the ``anisotropy'' force on the spin at site $i$ due to all neighbours $j$ within the cut-off of Eq. (\ref{eq8}) is:

\begin{linenomath*}
\begin{equation}
\begin{gathered}
\vec{f}_{{\rm anis},i} = \frac{1}{2} 
C_2 \sum _{j \neq i} 
\frac{1}{r_{\mathit{ij}}} 
\left[ \frac{\mathit{dg}\left( r_{\mathit{ij}}\right) }{dr_{\mathit{ij}}} 
\left\{  \vec{S_i}{\cdot}\vec{ S_i} + \vec{ S_j}{\cdot}\vec{
	S_j} \right\} \vec{r}_{\mathit{ij}} \right.  \\
	-\frac{\mathit{dg}\left(r_{\mathit{ij}}\right)}{dr_{\mathit{ij}}}\left\{\frac{\left(\vec{ r}_{\mathit{ij}}{\cdot}\vec{
		S_i}\right)^2}{r_{\mathit{ij}}^2}-\frac{\left(\vec{r}_{\mathit{ij}}\cdot \vec{
		S_j}\right)^2}{r_{\mathit{ij}}^2}\right\}\vec{r}_{\mathit{ij}}\\
		+2\frac{\mathit{dg}\left(r_{\mathit{ij}}\right)}{dr_{\mathit{ij}}}\left\{\left(\vec{
	r}_{\mathit{ij}}\cdot \vec{ S_i}\right)\vec{ S_i} +\left(\vec{
	r}_{\mathit{ij}}\cdot \vec{ S_j}\right)\vec{ S_j}\right\} \\ \left. +\frac{
1}{r_{\mathit{ij}}} \frac{d^2g\left(r_{\mathit{ij}}\right)}{d^2r_{\mathit{ij}}}\left\{\left(\vec{
	r}_{\mathit{ij}}{\cdot}\vec{ S_i}\right)^2 +\left(\vec{
	r}_{\mathit{ij}}{\cdot}\vec{ S_j}\right)^2\right\}\vec{
	r}_{\mathit{ij}}
\right] \label{eq9a}
\end{gathered}	
\end{equation}
\end{linenomath*}

Then, by substituting Eq.~(\ref{eq9a}) into (\ref{eq4}), and considering only the pair of particles $i$ and $j$, we obtain after some algebra,
\begin{linenomath*}
\begin{equation}
\begin{gathered}\left. \frac{d \vec{\lambda}_i}{dt} \right|_{j}+ \left.\frac{d \vec{\lambda}_j}{dt}\right|_{i} = \frac{1}{\hbar}\vec{r}_{ij} \times \vec{f}_{ji} \\
= \frac{C_2}{\hbar r_{\mathit{ij}}}\frac{\mathit{dg}\left(r_{\mathit{ij}}\right)}{dr_{\mathit{ij}}}\left[\left(\vec{
	r}_{\mathit{ij}}\cdot \vec{ S}_i\right)\vec{r}_{ij} \times \vec{S}_i +\left(\vec{
	r}_{\mathit{ij}}\cdot \vec{ S}_j\right)\vec{r}_{ij} \times \vec{ S}_j\right],\end{gathered}
\label{eq10}	
\end{equation}
\end{linenomath*}
where we have used the fact that the cross product of a vector with itself is zero. 

Clearly, Eqs.~(\ref{eq11}) and (\ref{eq10}) compensate each other when the cross products are commuted in Eq.~(\ref{eq10}) to appear in the same order as in Eq.~(\ref{eq11}). We have therefore shown analytically that the model of Ref. \cite{Perera2016} permits exchange of angular momentum, while at the same time guaranteeing conservation of total angular momentum.

\section{Numerical simulation results}

We validate our implementation of the anisotropy correction by studying the phenomenon known as the Einstein--de Haas effect \cite{EdH1915}. In its simplest version, the EdH experiment involves using a strong magnetic field to change the magnetization of a small freely suspended cylinder made of a ferromagnetic material. The sudden change in magnetization in the cylinder brought on by the strong magnetic field then manifests as a physical rotation about the axis of the sample as it evolves towards a new equilibrium magnetization due to the conservation of the total (spin + rotational) angular momentum.
The effect is well-known, but its microscopic origin has only been recently demonstrated in ultrafast (femtosecond scale) laser-induced demagnetization experiments \cite{Dornes2019}, though in that particular case the electronic degrees of freedom are also involved since the laser primarily excites the electrons \cite{Beaurepaire1996,Ma2012}.

\subsection{Simulation parameters for Fe}

In our numerical simulations of the EdH effect, we use a small prolate spheroid made of iron. We choose this shape because the anisotropy is uniaxial and a positive value of $C_2$ will see the spins aligning preferentially along the long axis of the spheroid. It consists of 505 Fe atoms cut out from a bulk periodic simulation cell of body-centered cubic (BCC) Fe(001). The particle is perfectly symmetric about its central axis along the $z$ direction, i.e., the $x$ and $y$ components of each of its constituent atoms sum to $0$ with respect to this axis. 

To describe the interatomic interactions, we use the Malerba et al. \cite{Malerba2010} EAM potential with a per-atom cohesive energy of $-4.122$ eV/atom. This contrasts with the $0$ eV cohesive energy of the Harmonic and Morse potentials used in Refs. \cite{Strungaru2021,Assmann2018}. The EAM potential used in this work is more realistic in that it is able to more accurately reproduce material properties \cite{Malerba2010}, including defects and the \textit{ab-initio} energy of a free Fe(001) surface.

Then, for the Heisenberg exchange term in Eq. (\ref{eq3}), we employ the same parameter values as in Table II of Ref. \cite{Strungaru2021}: $J_0 = 0.904$ eV and $r_{c} = 3.75$ \AA.

Finally, for the anisotropy term in Eq. (\ref{eq7}), we use $C_2 = 0.025$ eV.\AA$^2$ and $r_{0} = 3.5$ \AA. \footnote{We have discovered that the units of $C_1$ and $C_2$ are quoted incorrectly in Ref. \cite{Perera2016}, since they should be eV.\AA \ and eV.\AA$^2$, respectively.} This value of $C_2$ has been estimated by determining the difference in magnetic anisotropy energy (MAE) of the relaxed spheroid, with spins and atoms frozen, compared to the MAE with spins perpendicular to the first configuration, resulting in $\Delta \text{MAE} \sim 0.2$ meV/atom, in good agreement with the value of $\sim 0.2$ meV/atom reported in Fig. 7 of Ref. \cite{Li2013} for a similar-sized Fe nanocluster using a tight-binding formalism. \footnote{We note that our implementation does not model the weak static background anisotropy of 3$d$ transition metals, which, for example, gives rise to their preferred bulk magnetic easy axes. Rather, it accounts for dynamic anisotropy due to local symmetry-breaking about a given lattice site that may dominate at sufficiently high temperatures approaching the Debye temperature of the metal \cite{Eisenbach2015}. Nevertheless, dynamic many-body effects, such as those used in this model, play an important role \cite{Mentink2019} in fast demagnetization processes in ferromagnetic materials that cannot be modelled by static background anisotropy alone.}   

\subsection{SLD simulation of the Einstein--de Haas effect}

We simulate the EdH effect by mirroring an experiment where an initially demagnetized sample (such as via heat shock) is allowed to relax to an equilibrium magnetized state, while observing the rotational angular momentum which is induced by spins flipping and precessing as the sample reaches its equilibrium magnetization.

\begin{figure}[!ht]
\centering
\includegraphics[trim={0 0 0 0},clip,width=0.48\textwidth]{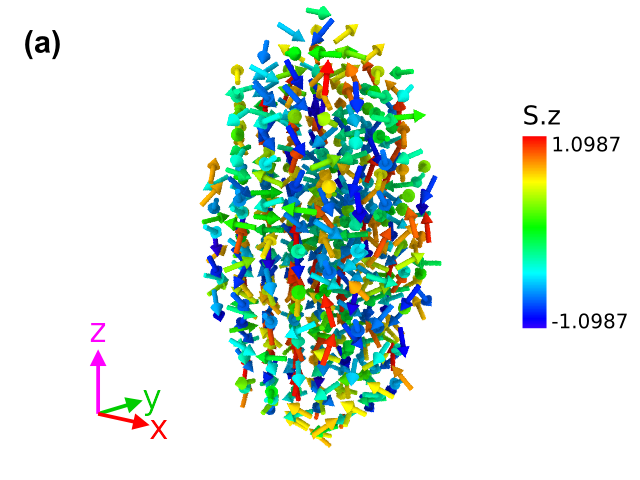}
\includegraphics[trim={0 0 0 0},clip,width=0.49\textwidth]{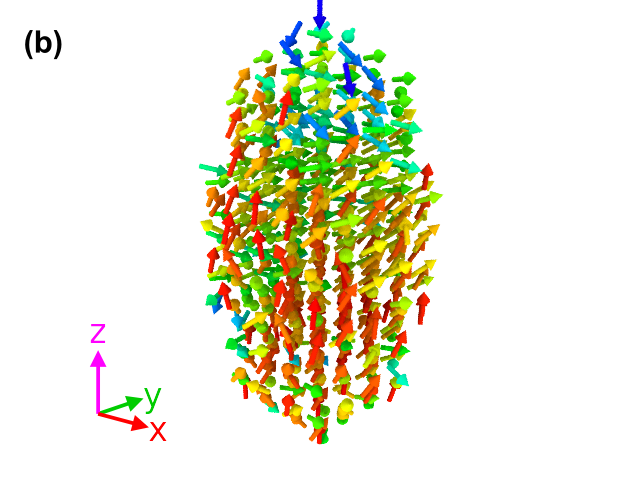}
\includegraphics[trim={0 0 2cm 2cm},clip,width=0.49\textwidth]{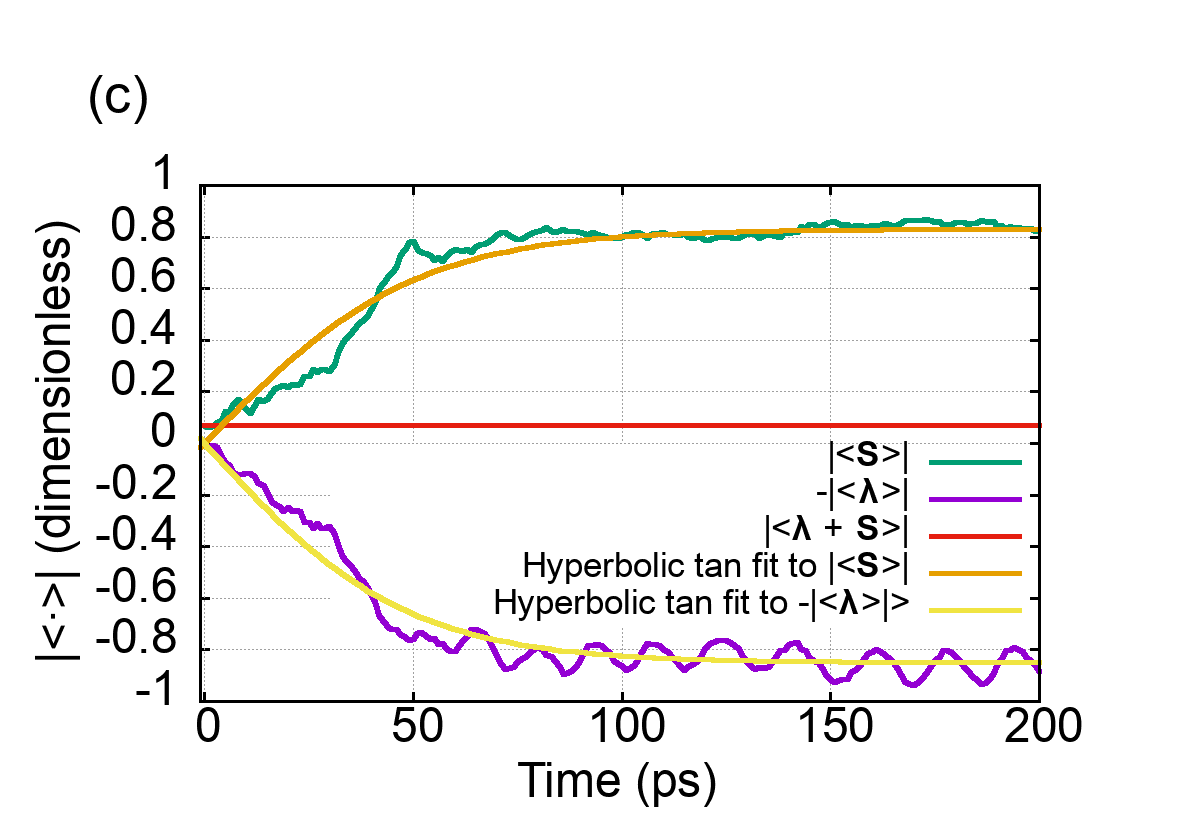}
\includegraphics[trim={0 0 2cm 2cm},clip,width=0.50\textwidth]{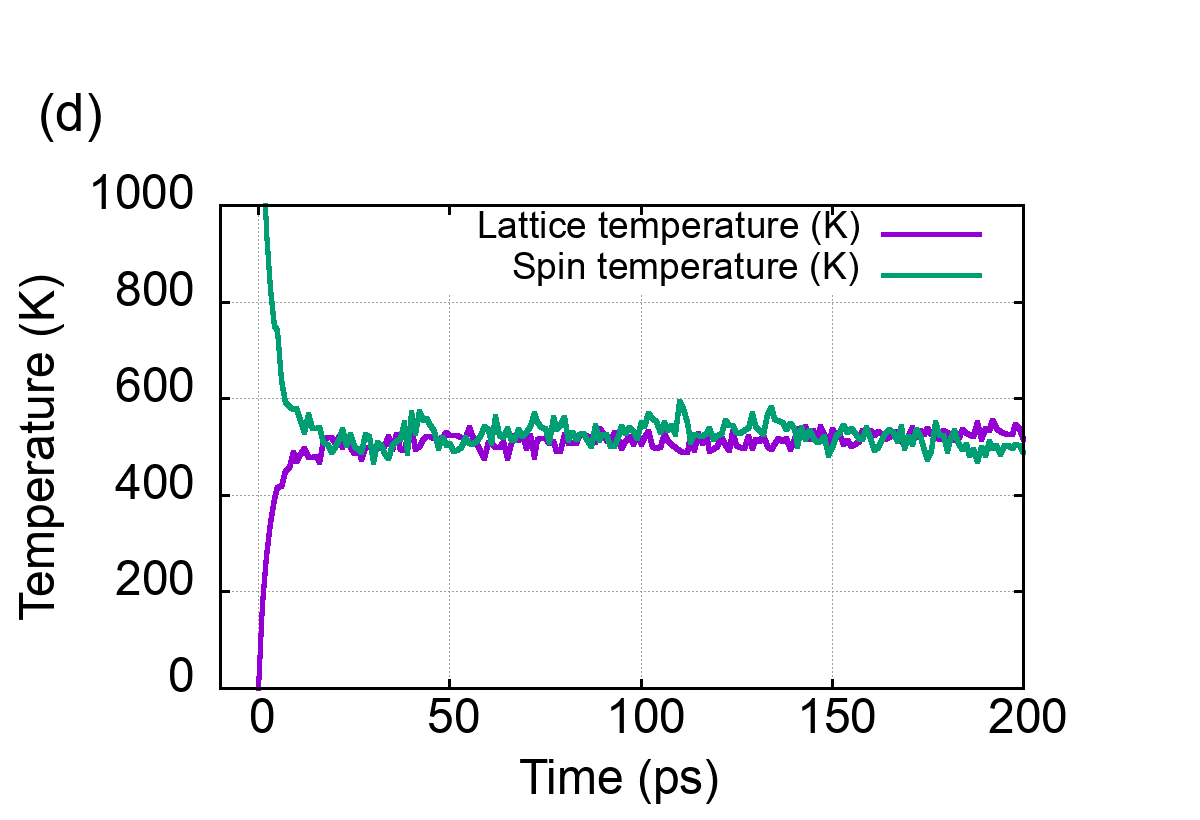}
\caption{(a) Initial ($t=0$) spin configuration of the Fe(001) prolate spheroid. The atoms (not shown) are placed near the sites they would occupy in a perfectly ordered BCC lattice of Fe(001) and the spins are initially randomly oriented. (b) After $t=100$ ps, the spins mostly point upward. The color legend refers to the magnitude of the (dimensionless) atomic spins on the positive \textit{z}-axis, i.e., the direction of saturation magnetization of Fe, $2.2 \mu_{\rm B}$, divided by the electron's gyromagnetic factor, $g=2.002319$, and $1.0 \mu_{\rm B}$. (c) Conservation of total average per-atom angular momentum (red) as a sum of contributions from the atomic spins $\left|\left<\vec{S}\right>\right|$ (green) and lattice $-\left|\left<\vec{\lambda}\right>\right|$ (purple) for $C_2=0.025$ eV.\AA$^2$. For convenience, the negative of $\left|\left<\vec{\lambda}\right>\right|$ has been plotted so that $\left|\left<\vec{S}\right>\right|$ coincides with the general orientation of $\left<\vec{S}\right>$ in (b). Hyperbolic tangent fits to $\left|\left<\vec{S}\right>\right|$ (orange) and $-\left|\left<\vec{\lambda}\right>\right|$ (yellow) are also shown. (d)  The lattice temperature (purple), initially $0$ K, rises rapidly while the spin temperature (green), initially $\sim 12000$ K, converges equally rapidly to the equilibrium temperature of $\sim 500$ K for $C_2=0.025$ eV.\AA$^2$.}
\label{fig1}
\end{figure}

We therefore use an input configuration for our SLD simulation which consists of fully randomized spins, as shown in Fig. \ref{fig1} (a). The sample is then allowed to relax for a period of $1$ ns, reaching an equilibrium magnetized state after $\sim 100$ ps, as illustrated in Fig. \ref{fig1} (b), where we see the spins have mostly aligned parallel to the easy axis. \footnote{See Supplemental Material at http://link.xxx.org/supplemental/xxxx/xxxx.xxx for movies of the atoms and spins of the spheroid in Fig. \ref{fig1} for $C_2 = 0.025, 0.05, 0.075, 0.1, 0.15$ and $0.2$ eV.\AA$^2$ showing how the nanocluster evolves to the final states depicted in Fig. \ref{fig1}}.  

In Fig. \ref{fig1} (c) we plot the time evolution of the average per-atom spin and rotational angular momentum for $C_2=0.025$ eV.\AA$^2$. Both the rotational lattice and spin angular momenta are initially near $0$. As the simulation proceeds, angular momentum is transferred from the spins as they rotate (during the process of reaching a final equilibrium magnetized state), to rotational lattice angular momentum, while conserving the total angular momentum. The magnitude of the total average angular momentum, or red line, in Fig. \ref{fig1} (c) has been calculated as

\begin{linenomath*}
\begin{equation}
\begin{gathered}
\left|\left<\vec{\lambda}+\vec{S}\right>\right| = \left({[\left <\lambda_{x}\right>+\left<S_{x}\right>]}^2+{[\left<\lambda_{y}\right>+\left <S_{y}\right>]}^2+{[\left<\lambda_{z}\right>+\left<S_{z}\right>]}^2 \right)^{1/2}.
\end{gathered}
\end{equation}
\end{linenomath*}

We note that the average per atom spin angular momentum of $\sim 0.8 \left|\vec S\right|_{\rm initial}$ is higher than expected at the temperature\footnote{\texttt{SPILADY} calculates the lattice temperature from the equipartition of kinetic energy in three dimensions and we implemented the generalized expression for the spin temperature developed by Nurdin et al. \cite{nurdin2000} in \texttt{SPILADY}.}  of $\sim 500$ K shown in Fig. \ref{fig1} (d), due to the model of Refs. \cite{Ma2008,Perera2016,Strungaru2021} which does not account for size dependence of magnetization in nanoparticles \cite{DosSantos2020,DosSantos2021}.

\begin{figure}[!ht]
\centering
\includegraphics[width=0.97\textwidth]{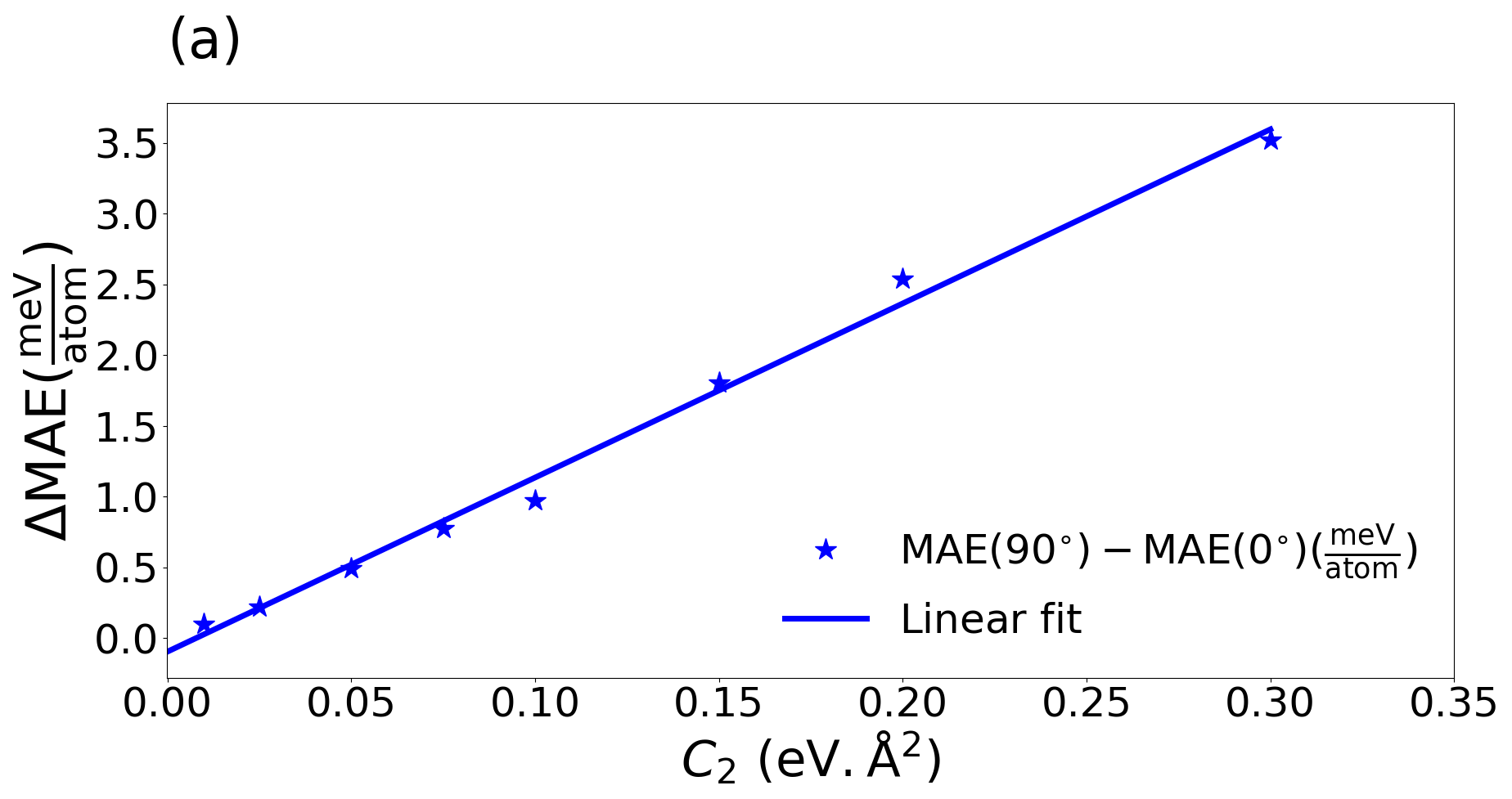}
\includegraphics[width=0.97\textwidth]{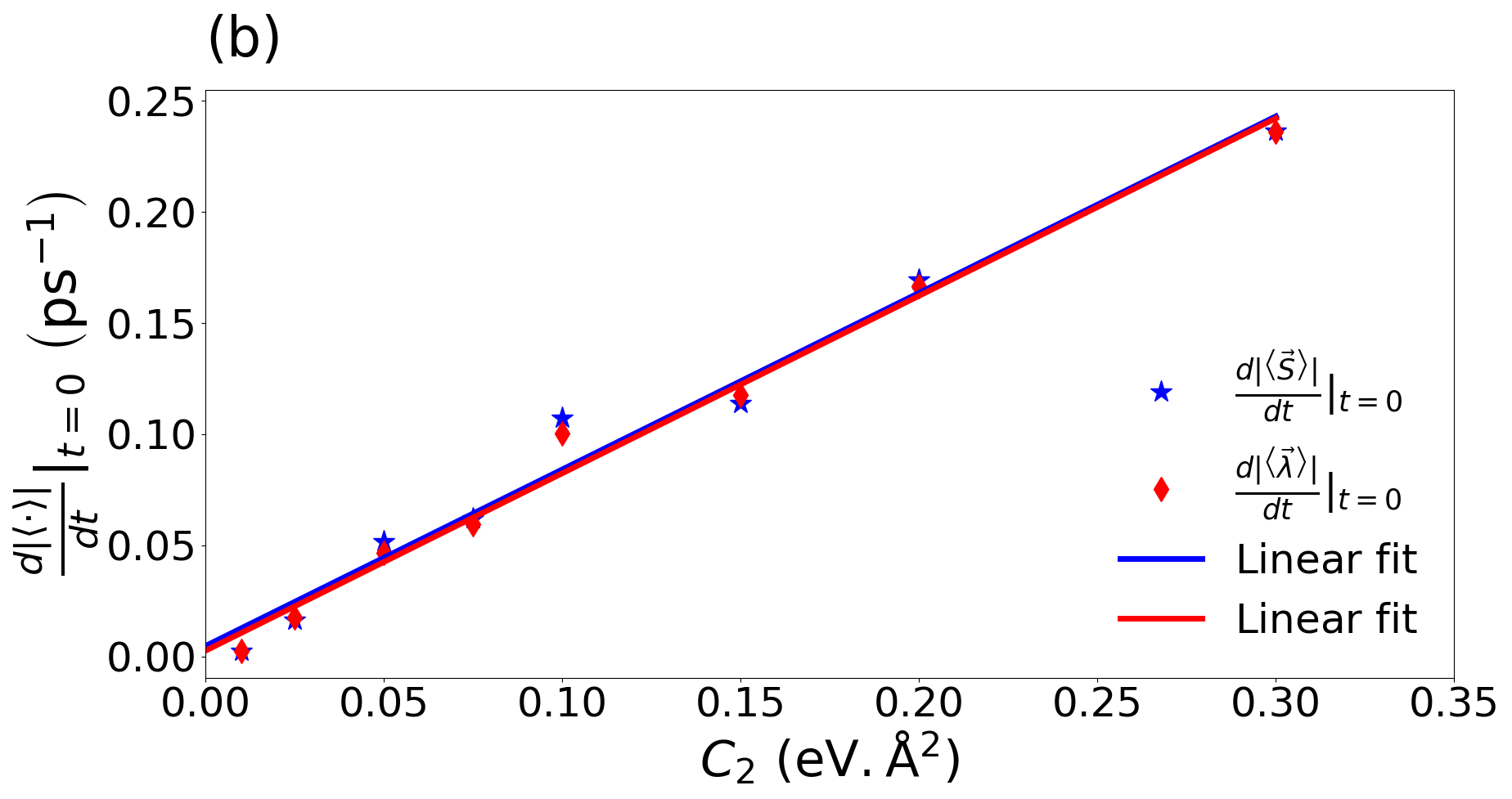}
\caption{(a) The magnetic anisotropy energy difference scales linearly with $C_2$ ($R^2 =0.993$) and the range of $\Delta \text{MAE}$ values are $\sim 1$ meV/atom \cite{Li2013}. (b) The dependence on $C_2$ of the rate of change in angular momentum is clear, with stronger anisotropic interactions (larger $C_2$) resulting in more rapid change in angular momenta. (Here, $R^2 =0.978$ in the case of the spins and $ 0.988$ in the case of the lattice.)}
\label{fig2}
\end{figure}

In Fig. \ref{fig2} (a) we plot the difference in magnetic anisotropy energy $\Delta \text{MAE}$ of the spin populations at $90$\textdegree \ to one another as a function of $C_2$; it is seen that $\Delta \text{MAE}$ is linear in $C_2$, verifying that the MAE is directly proportional to $C_2$.

Further, we verify that the strength of the anisotropy correction $C_2$ determines the rate of exchange of angular momentum between spins and lattice rotational angular momentum. To this end, and for ease of comparison, we fit the absolute values of the average per-atom lattice and spin angular momenta to the hyperbolic tangent function $a\tanh\left(bt\right)$ using the first $500$ ps of simulation data in each case, as illustrated in Fig. \ref{fig1} (c) for the particular case of $C_2=0.025$ eV.\AA$^2$. The rate of change of angular momentum is the greatest near the origin; therefore we use the derivatives of the fits to this function at $t=0$ as measure of the rate of exchange of angular momentum. In Fig. \ref{fig2} (b) we plot the rates of change of spin and lattice rotational angular momenta as a function of $C_2$. It is seen that the rate of exchange of angular momenta is linear in $C_2$, verifying that this rate is directly proportional to the strength of the magnetic anisotropy.

\begin{figure}[!ht]
\centering
\includegraphics[width=0.48\textwidth]{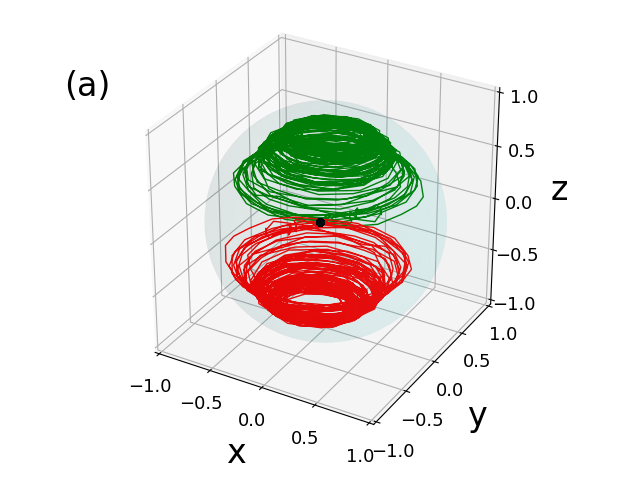}
\includegraphics[width=0.48\textwidth]{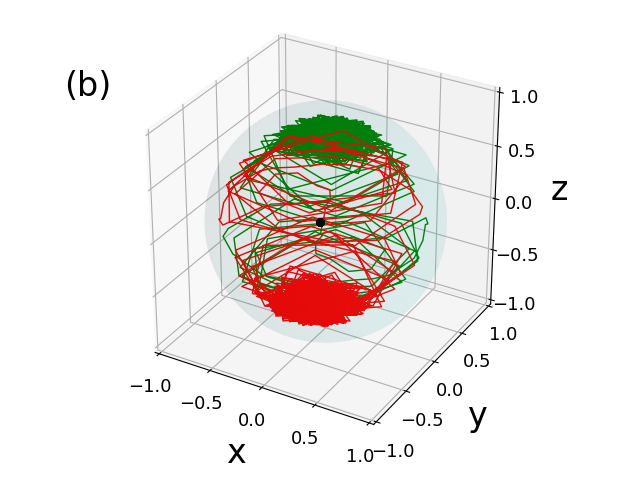}
\includegraphics[trim={24cm 10cm 0 10cm},clip,width=0.78\textwidth]{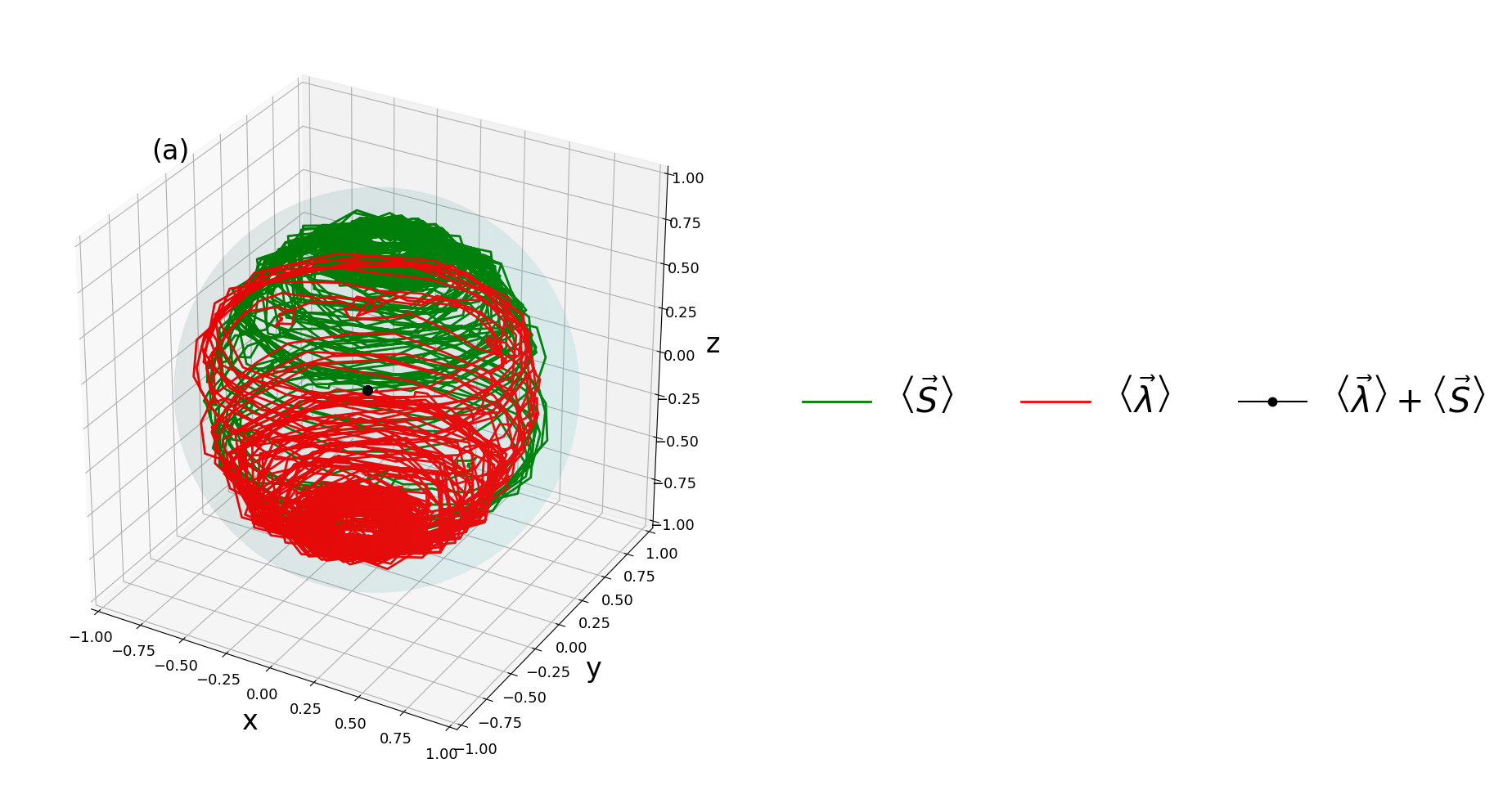}
\includegraphics[width=0.48\textwidth]{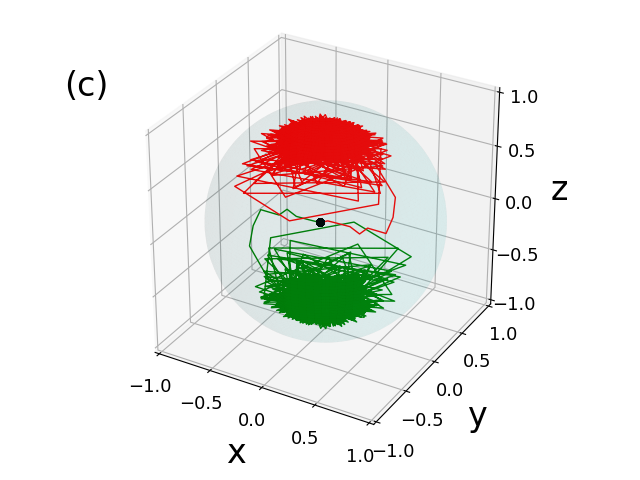}
\includegraphics[width=0.48\textwidth]{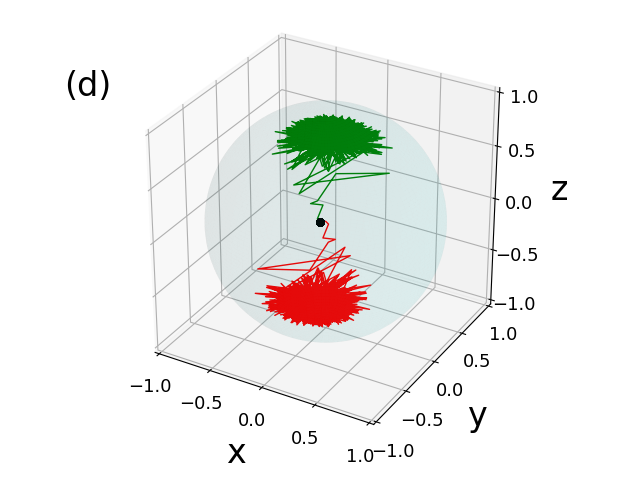}
\caption{Conservation of total angular momentum for the full simulation duration and different values of the anisotropy coefficient $C_2$: (a) $0.025$, (b) $0.075$, (c) $0.1$, (d) $0.15$ eV.\AA$^2$. Total angular momentum (black) does not vary during the simulation, while the individual components, the average per atom spin (green curve) and lattice rotational moment (red curve) do vary.  As the value of $C_2$ increases, the coordinates of the average per-atom spin and lattice angular momenta take less time to approach the long ($z$) axis of the spheroid via exchange of angular momentum. This is also evident from the values of the derivatives at $t=0$ of the hyperbolic tangent fits to the absolute values of the spin and lattice angular momenta in Fig. \ref{fig2}. }
\label{fig3}
\end{figure}

In order to provide visual confirmation of the trend in Fig. \ref{fig2} (b), we plot in Fig.~\ref{fig3} the time evolution of the coordinates of the per-atom average angular momenta of the spins (green), lattice (red) and their sum (black) for a range of values of $C_2$. Since the anisotropy is uniaxial in this work, the green and red traces tend to cluster near the $z$-axis, and with opposite signs. The symmetry of the spin Hamiltonian implies that there is no preferred final direction of the average per-atom spin and lattice angular momenta as long as they are mutually opposite and generally point along the cluster's easy axis ($z$ in this case). Therefore, randomly, either of the spin or the lattice rotational angular momentum can be oriented along the positive $z$ axis (see e.g. the traces of Fig. \ref{fig3} (c) vs the other three traces); it is the relative orientation of rotational versus spin angular momenta which is important, and which are always opposite and equal, conserving total angular momentum.

Finally, we note that, to our knowledge, the only previous SLD simulation we are aware of that reproduces the EdH effect (Ref. \cite{Assmann2018}) applies an external magnetic field of $50$ Tesla throughout the entire demagnetization process of a free face-centered cubic Co cluster. Also, only the $z$ components of the spin and lattice angular momenta are plotted in Ref. \cite{Assmann2018}, casting doubt on whether \textit{total} angular momentum is actually conserved in their EdH simulation. (An external field can induce a non-zero overall spin torque due to individual contributions not cancelling out in a pairwise manner.)  

\section{Conclusion}

We have shown analytically that the dynamic many-body model of uniaxial anisotropy described in Ref. \cite{Perera2016} conserves the total angular momentum and energy of the system. We also implemented this model in the spin-lattice dynamics code \texttt{SPILADY}~\cite{spilady2016}, and also make the source code available online \cite{SPILADYSOC}. Our implementation reproduces the full Einstein--de Haas effect in an isolated iron nanoparticle, consistent with the conservation of total angular momentum by the model over a range of values of the magnetic anisotropy coefficient $C_2$. The value of this coefficient not only determines the rate at which exchange between spin and rotational angular momenta occurs, but also how quickly the angular momenta align parallel to the long axis of the nanoparticle for positive $C_2$. 

The advantage of the approach followed in this work, compared to other similar generally applicable semi-classical models, is that the realistic EAM potential is used to describe interactions between the atoms, allowing accurate reproduction of \textit{ab-initio} free surface energies and defects. Another advantage is that the angular momentum exchange occurs in a very stable and consistent manner, with full spin-lattice relaxation occurring in agreement with previously reported times $\sim 100$ ps. Moreover, neglecting shape anisotropy, the difference in per-atom magnetic anisotropy energy between spin populations oriented at $90$\textdegree \ to each other in the same frozen atomic configuration, yields order-of-magnitude agreement, at the lower range of the magnetic anisotropy interaction strength, $C_2$, with previously reported tight-binding values for Fe nanoclusters of comparable size. The development of dynamic anisotropy models due to local symmetry breaking at finite temperatures is therefore crucial to modeling fast demagnetization processes.

Our implementation of anisotropy can be employed to study any single-element ferromagnetic system in which local coupling between spins and atoms cannot be ignored. It is also possible to extend this approach to multiple-element systems once \texttt{SPILADY}, the underlying spin-lattice dynamics code used in this implementation, makes provision for this capability. For this reason it is likely to find applications in a significant number of situations where uniaxial magnetization is important and coupling between lattice and spins plays a non-negligible role.

\section*{Acknowledgements}

We acknowledge financial support from the Ministry of Science and Innovation of Spain (grant No. PID2019-109539GB-4). This work was supported by the Generalitat Valenciana through PROMETEO2017/139 and PROMETEO/2021/017. C. S. gratefully acknowledges financial supports from Generalitat Valenciana with (CDEIGENT2018/028). JFR acknowledges funding from FCT grant  PTDC/FIS-MAC/2045/2021. The SLD calculations in this paper were performed on the high-performance computing facilities of the University of South Africa.







\end{document}